\documentclass[twocolumn,groupedaddress,amsmath,amssymb,prl,aps,longbibliography]{revtex4-1}

\usepackage{graphicx}% Include figure files
\usepackage{dcolumn}% Align table columns on decimal point
\usepackage{bm}% bold math
\usepackage{soul} %strikethrough - MEG
\usepackage{xcolor} % color text - MEG
\usepackage{amsmath} % more tools for math environment - MEG

\begin{document}

\title{Extrinsic nature of the polarization in hafnia ferroelectrics}

\author{Binayak Mukherjee,$^{1}$ Natalya S. Fedorova,$^{1}$ and Jorge
  \'I\~niguez-Gonz\'alez$^{1,2}$}

\affiliation{
  \mbox{$^{1}$Smart Materials Unit, Luxembourg Institute of Science
    and Technology (LIST),} \mbox{Avenue des Hauts-Fourneaux 5, L-4362
    Esch/Alzette, Luxembourg}\\
 \mbox{$^{2}$Department of Physics and Materials Science, University
   of Luxembourg, Rue du Brill 41, L-4422 Belvaux, Luxembourg}}

\begin{abstract}
Hafnia and related fluorites defy our understanding of
ferroelectricity, even if we restrict ourselves to the intrinsic
properties of ideal crystals. Here we focus on a critical puzzle,
namely, the sign of the electric polarization. Using first-principles
simulations, we show that a polar hafnia layer with a fixed atomic
configuration can give rise to depolarizing fields of either positive
or negative sign, depending on the environment. This implies that (the
sign of) the polarization in hafnia is extrinsic in nature. We explain
this result and discuss its relevance to other ferroelectric families.
\end{abstract}

\maketitle   

Ferroelectric HfO$_{2}$ has attracted considerable attention, to a
large extent because of its industrial significance as a
silicon-compatible ferroelectric~\cite{silva23b}. Additionally, and
somewhat surprisingly, understanding the nature of ferroelectricity in
HfO$_{2}$, in contrast to archetypal perovskites like BaTiO$_{3}$ or
PbTiO$_{3}$, is proving to be an ongoing
challenge~\cite{iniguez-gonzalez24}.

Unlike in the perovskites, the ferroelectric phase of HfO$_{2}$
(usually labeled oIII, with space group $Pca2_{1}$) is not the ground
state, but rather a metastable polymorph that can be obtained in thin
films~\cite{schroeder22}. Furthermore, in hafnia there is no obvious
paraelectric phase whose lattice instabilities give rise to the oIII
structure~\cite{aramberri23,iniguez-gonzalez24}. Indeed, the non-polar
polymorphs experimentally observed in the relevant temperature range
-- i.e., the tetragonal ($P4_{2}/nmc$) and monoclinic ($P2_{1}/c$)
phases -- are local minima of the energy~\cite{fan22}. Alternative
reference structures have been proposed
theoretically~\cite{zhou22b,aramberri23,raeliarijaona23,jung25,qi25},
but there is no experimental evidence of their relevance concerning
the occurrence or properties of the oIII phase. In
Ref.~\onlinecite{iniguez-gonzalez24}, some of us have discussed the
absence of a high-symmetry reference phase for hafnia and the
consequences of this state of matters.

\begin{figure*}
    \centering
    \includegraphics[width=0.8\linewidth]{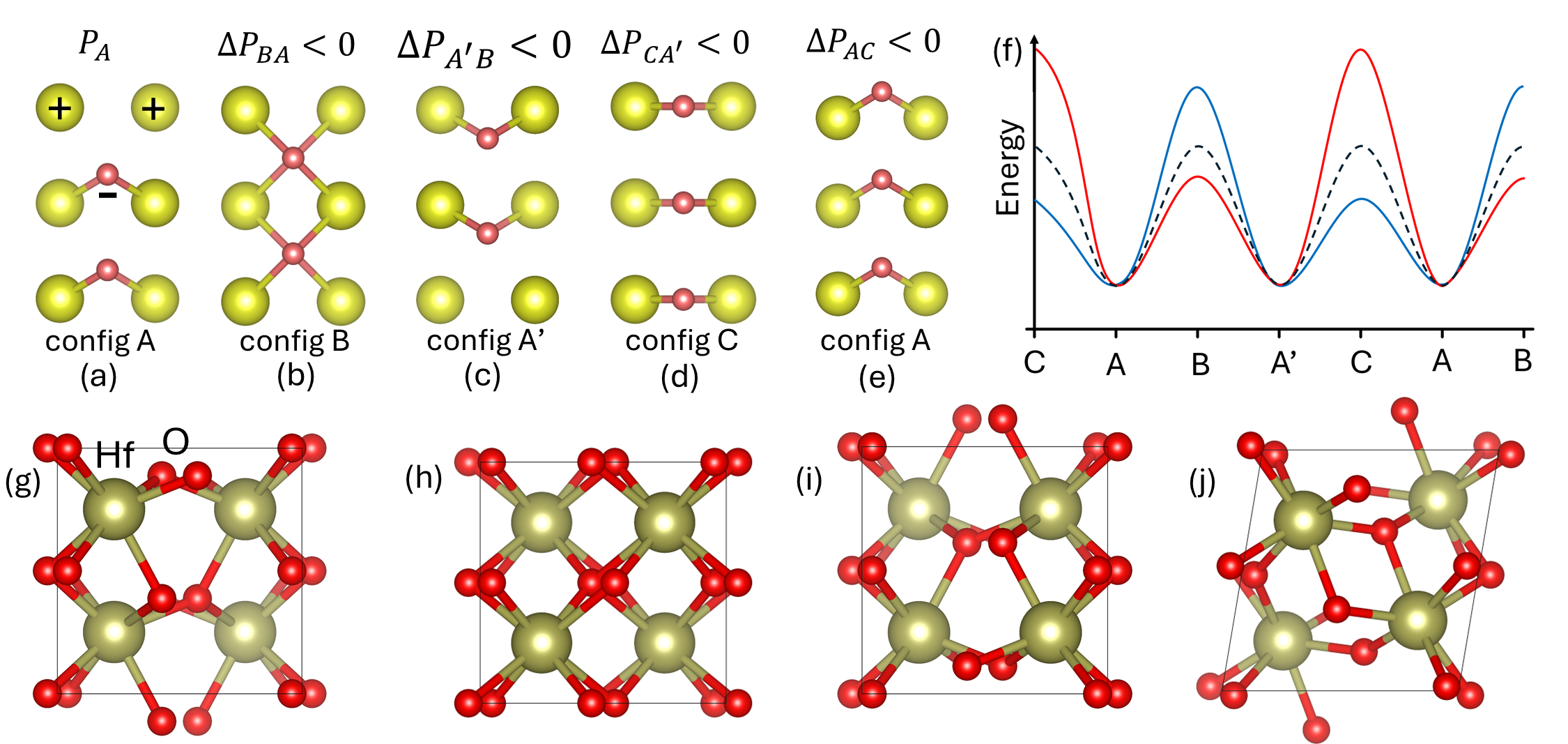}
    \caption{Panels~(a) to (f): Hypothetical polar material displaying
      multiple centrosymmetric reference states. Starting from a polar
      configuration~A (a), the red anion moves upward to yield a
      centrosymmetric configuration B (b), then to a state A$'$ with a
      polarization reversed with respect to that of A (c), followed by
      a different centrosymmetric configuration~C (d), and finally
      back to state A (e). Panel~(f) shows three possible energy
      landscapes connecting these states. Panels~(g) to (j) show
      hafnia polymorphs we can identify with these idealized
      configurations: configuration~A would be the oIII phase of
      hafnia in a polar state we denote oIII-1 (g), B would be the
      centrosymmetric tetragonal polymorph (h), A$'$ would be the
      state labeled oIII-2 with polarization opposite to that of
      oIII-1 (i), and C would be the centrosymmetric monoclinic ground
      state.}
    \label{fig:schematic}
\end{figure*}

Here we focus on an aspect that we consider even more important, at
both fundamental and practical levels: the nature of the polarization
in the oIII phase~\cite{qi22,aramberri23}. To introduce the problem in
a general way, we resort to the sketch in
Fig.~\ref{fig:schematic}. Suppose we have a material with two ions in
the unit cell, one positively and one negatively charged. We start
from the polar configuration~A (Fig.~\ref{fig:schematic}(a)) and move
the anion upwards until we reach the centrosymmetric configuration~B
(Fig.~\ref{fig:schematic}(b)). A further upward shift of the anion
leads us to~A$'$ (Fig.~\ref{fig:schematic}(c)), equivalent by symmetry
to A but oppositely polarized. Another upward displacement renders a
second centrosymmetric configuration~C (Fig.~\ref{fig:schematic}(d)),
before we reach a state that is identical to the starting
configuration~A (Fig.~\ref{fig:schematic}(e)). All the polarization
changes along this path are negative – e.g., $\Delta P_{BA} = P_{B} -
P_{A} < 0$.

Now we can ask: what is the polarization of configuration~A? Suppose
that the energy landscape is given by the red curve in
Fig.~\ref{fig:schematic}(f): configuration~B has a relatively low
energy and provides us with a transition state, while C is very high
in energy and thus inaccessible. In such conditions, applying a
negative electric field along the vertical direction to the system in
configuration~A will cause a transition -- through~B --
to~A$'$. Hence, we have $P_{A} = -(\Delta P_{\rm A'B}+\Delta P_{\rm
  BA})/2 > 0$. By contrast, if we apply a positive field to A, we will
simply cause the polarization to saturate, as C is not accessible.

However, if the energy landscape is given by the blue curve in
Fig.~\ref{fig:schematic}(f), then C is accessible and B is not. In
this case, switching from A to A$'$ can only proceed through C – by
applying a positive electric field to A – and we have $P_{A} =
-(\Delta P_{\rm A'C}+\Delta P_{\rm CA})/2 < 0$. Thus, the physically
available switching path determines the sign of the polarization. A
third scenario is given by the black curve in
Fig.~\ref{fig:schematic}(f), where both switching paths -- through B
or C -- are energetically available. This scenario seems to describe
HfO$_{2}$ most closely. As shown in Figs.~\ref{fig:schematic}(g) to
\ref{fig:schematic}(j), well-known polymorphs of hafnia can be readily
associated to the configurations of our idealized crystal. (The
so-called ``active oxygens'' in hafnia -- which can be viewed as
responsible for the electric polarization of the oIII phase -- would
correspond to the mobile anions in our model crystal.) Further,
growing simulation evidence suggests that switching can happen through
these (and other) states, rendering multiple competitive and
non-equivalent
pathways~\cite{clima14,choe21,qi22,silva23a,ma23,wu23,dou24,hu25}. We
thus have that, for a given polar state of hafnia – e.g., oIII-1 as
defined in Fig.~\ref{fig:schematic}(g) --, we cannot tell if the
polarization points up or down.

One may wonder whether this is a theoretical subtlety or if it has
tangible physical consequences. For example, a hafnia sample in the
$+P$ polarization state can be expected to give rise to a negative
depolarizing field. But then, what is the sign of the depolarizing
field corresponding to state oIII-1, if the sign of the polarization
cannot be determined from the atomic configuration alone? We arrived
at this conundrum (and its solution) serendipitously, while
investigating the electrostatics of HfO$_{2}$/GeO$_{2}$ polar/nonpolar
superlattices (SLs) with polarization along the stacking direction,
which we had previously predicted to be energetically
viable~\cite{mukherjee24}. In such SLs, the depolarizing field --
arising as a reaction to the polar distortion of the HfO$_{2}$ layer
-- overlaps with the band bending typical of any such heterostructure,
which complicates the analysis. Hence, to isolate the depolarizing
effects, here we simulate pure HfO$_{2}$/HfO$_{2}$ SLs where one layer
is in a polar configuration (always oIII-1 as specified in
Fig.~\ref{fig:schematic}(g), for concreteness) while the other is in a
non-polar state. In the following we summarize our findings. (See
Methods for details of the simulations.)

\begin{figure*}
    \centering
    \includegraphics[width=0.8\linewidth]{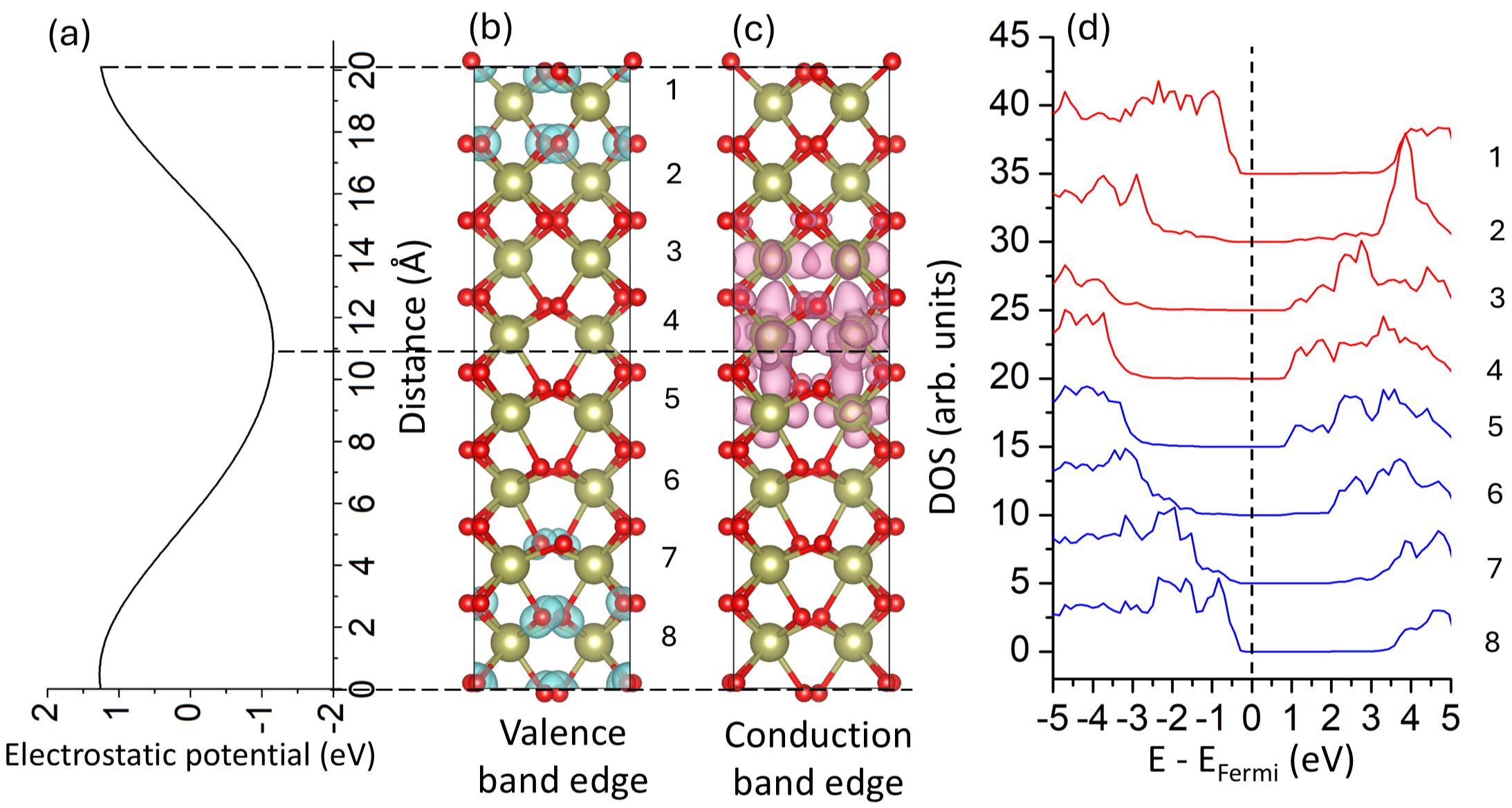}
    \caption{Results for the oIII-1/tetragonal superlattice. Panel~(a)
      shows the macroscopically-averaged electrostatic potential; (b)
      shows the structure of the superlattice superposed with the
      charge density of occupied states lying within 2~eV below the
      Fermi level; (c) shows the charge density of unoccupied states
      lying within 2~eV above the Fermi level; (d) shows the
      layer-resolved electronic density of states. Dashed lines in
      panels~(a) to (c) indicate the location of the interfaces
      between oIII-1 and tetragonal layers.}
    \label{fig:t-oIII}
\end{figure*}

Let us start by discussing in detail the representative case of a
superlattice combining layers in the polar oIII-1 and non-polar
tetragonal configurations. Our results are summarized in
Fig.~\ref{fig:t-oIII}, where panels~(b) and (c) show the atomic
structure. The Hf-planes numbered 1 to 3 are in the tetragonal state,
while planes 5 to 7 display the oIII-1 configuration. The interfacial
regions -- roughly corresponding to Hf-planes~4 and 8, where the atoms
are allowed to relax -- present a distorted structure that differs
from the ideal polymorphs. Fig.~\ref{fig:t-oIII}(d) shows the
plane-resolved electronic density of states. We observe a band bending
such that the interfaces at Hf-planes~4 and 8 would attract extra
electrons and extra holes, respectively. This is consistent with the
results for the macroscopically-averaged electrostatic potential
(Fig.~\ref{fig:t-oIII}(a), where the sign is chosen so that potential
minima attract electrons) and with the spatial location of the states
around the top of the valence band (Fig.~\ref{fig:t-oIII}(b)) and
around the bottom of the conduction band
(Fig.~\ref{fig:t-oIII}(c)). Indeed, given that the band offset between
these two hafnia polymorphs is relatively small (Supp. Fig.~1), and
that the simulated SL remains insulating, these results can be safely
interpreted as being consistent with a negative depolarizing field
within the oIII-1 layer, so that any excess negative (resp. positive)
charges would travel to the interface around Hf-plane number~4
(resp.~8). To test this picture, we run simulations with an
exaggerated smearing energy -- which effectively provides us with an
excess of electrons and holes -- and confirmed the expected spatial
distribution of the screening charges (Supp. Fig.~2). Thus, according
to these results, we would say that the simulated oIII-1 layer is
subject to a negative depolarizing field, which suggests that its
electric polarization is positive.

\begin{figure*}
    \centering
    \includegraphics[width=0.8\linewidth]{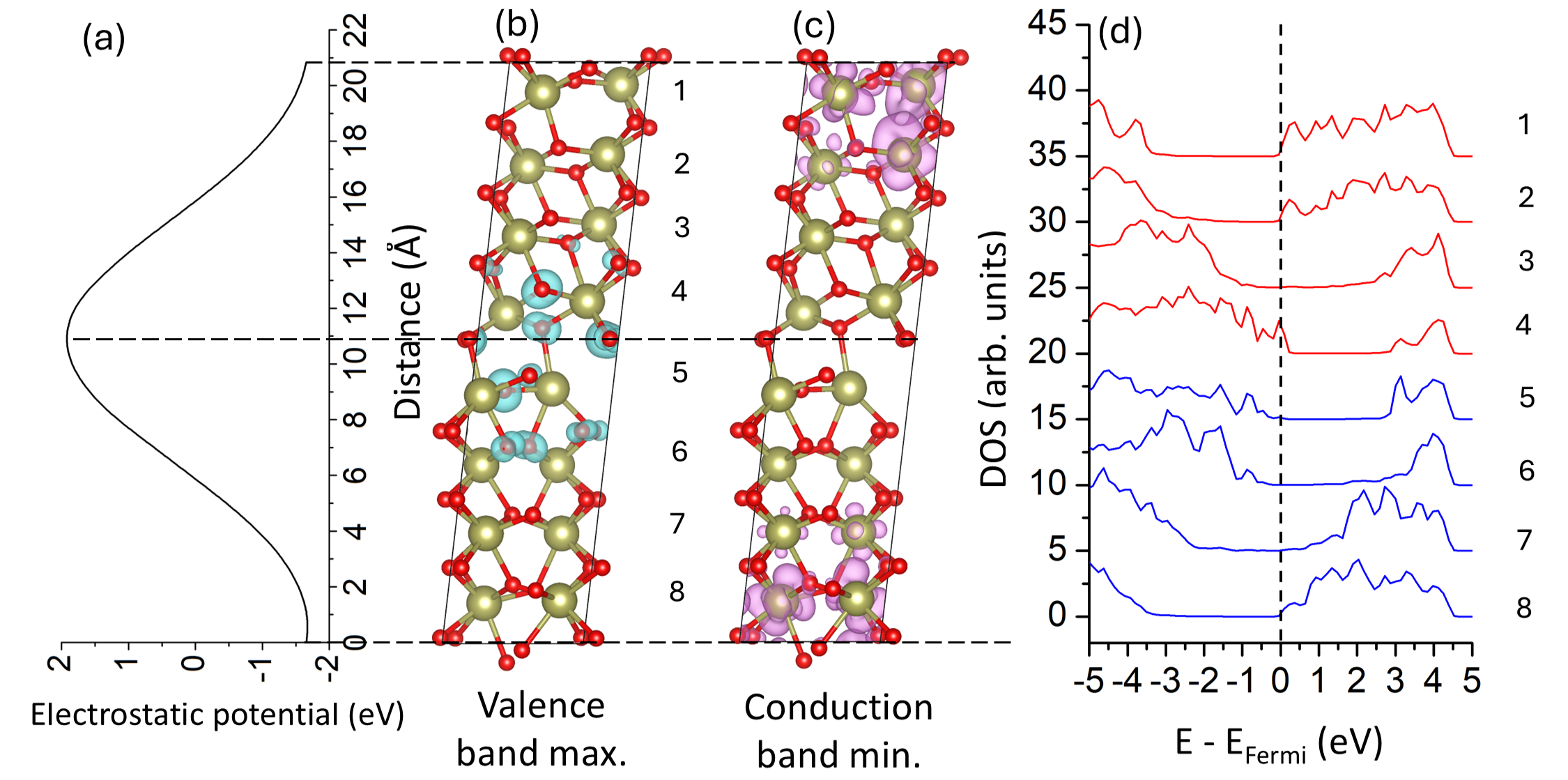}
    \caption{Same as Fig.~\protect\ref{fig:t-oIII}, but for the
      oIII-1/monoclinic all-hafnia superlattice.}
    \label{fig:m-oIII}
\end{figure*}

Consider now a superlattice that combines a polar layer in the oIII-1
state (exactly as before) with a non-polar layer that has the
structure of hafnia's monoclinic ground state. Common wisdom would
suggest that the results for this second SL should be similar to those
just presented for the oIII-1/tetragonal system, since the situations
seem qualitatively identical. It is therefore remarkable that, as
shown in Fig.~\ref{fig:m-oIII}, our calculations reveal a fully
reversed scenario. We find that, for the oIII-1/monoclinic
superlattice, the oIII-1 layer is subject to a positive depolarizing
field, as shown by the computed electrostatic potential
(Fig.~\ref{fig:m-oIII}(a)) and band bending
(Fig.~\ref{fig:m-oIII}(d)), suggesting that excess negative
(resp. positive) charges will flow to the interface around Hf-plane
number~8 (resp.~4). As before, we confirm this interpretation by
running a simulation in which we artificially create excess electrons
and holes, which flow to the interfaces corresponding to
conduction-band bottom (Fig.~\ref{fig:m-oIII}(c)) and valence-band top
(Fig.~\ref{fig:m-oIII}(b)), respectively (see Supp. Fig.~3). Thus, in
this SL, the oIII-1 layer is subject to a positive depolarizing field,
which suggests that its electric polarization is negative.

Interestingly, the above observations lend themselves to a simple
explanation once we recall the arbitrariness in the sign of the
polarization of oIII hafnia, discussed above. Imagine a homogeneous
tetragonal structure, which we distort to obtain the SL state depicted
in Fig.~\ref{fig:t-oIII}. In essence, this will involve a downward
displacement of the active oxygens, which will result in a positive
polarization change. Such a positive $\Delta P$ is consistent with the
negative depolarizing field revealed by the results in
Fig.~\ref{fig:t-oIII}. Similarly, if we start from a homogeneous
monoclinic state and distort the structure to obtain the configuration
in Fig.~\ref{fig:m-oIII}, we will be essentially moving active oxygens
upwards and inducing a negative polarization change, which is
consistent with the results in Fig.~\ref{fig:m-oIII}. In Supp. Fig.~4
we provide pictorial representations of the mentioned distortions and
induced polarizations.

Indeed, if we follow the movement of oxygen anions relative to the
planes of hafnium cations, we find that the tetragonal phase of hafnia
can be readily identified with configuration B in
Fig.~\ref{fig:schematic}, while the monoclinic state is a natural
choice for configuration C. But then, interestingly, hafnia's cubic
polymorph (shown in Supp. Fig.~5) would seem to be another possible
choice for configuration B, while the oVII polymorph depicted in
Supp. Fig.~6 would be a reasonable choice for configuration C. Indeed,
our calculations confirm that the oIII-1/cubic superlattice displays a
negative depolarizing field (Supp. Fig.~5(b)), exactly as the
oIII-1/tetragonal case (Fig.~\ref{fig:t-oIII}(a)); correspondingly, we
also find that the results for the oIII-1/oVII superlattice
(Supp. Fig.~6(b)) closely resemble those of the oIII-1/monoclinic
system (Fig.~\ref{fig:m-oIII}(a)). Hence, a picture emerges whereby
the polarization of the oIII-1 layer is consistent with a
centrosymmetric reference that is effectively defined by the
surrounding materials. Remarkably, this implies that the sign of the
polarization is extrinsic in nature.

\begin{figure}
    \centering
    \includegraphics[width=0.8\linewidth]{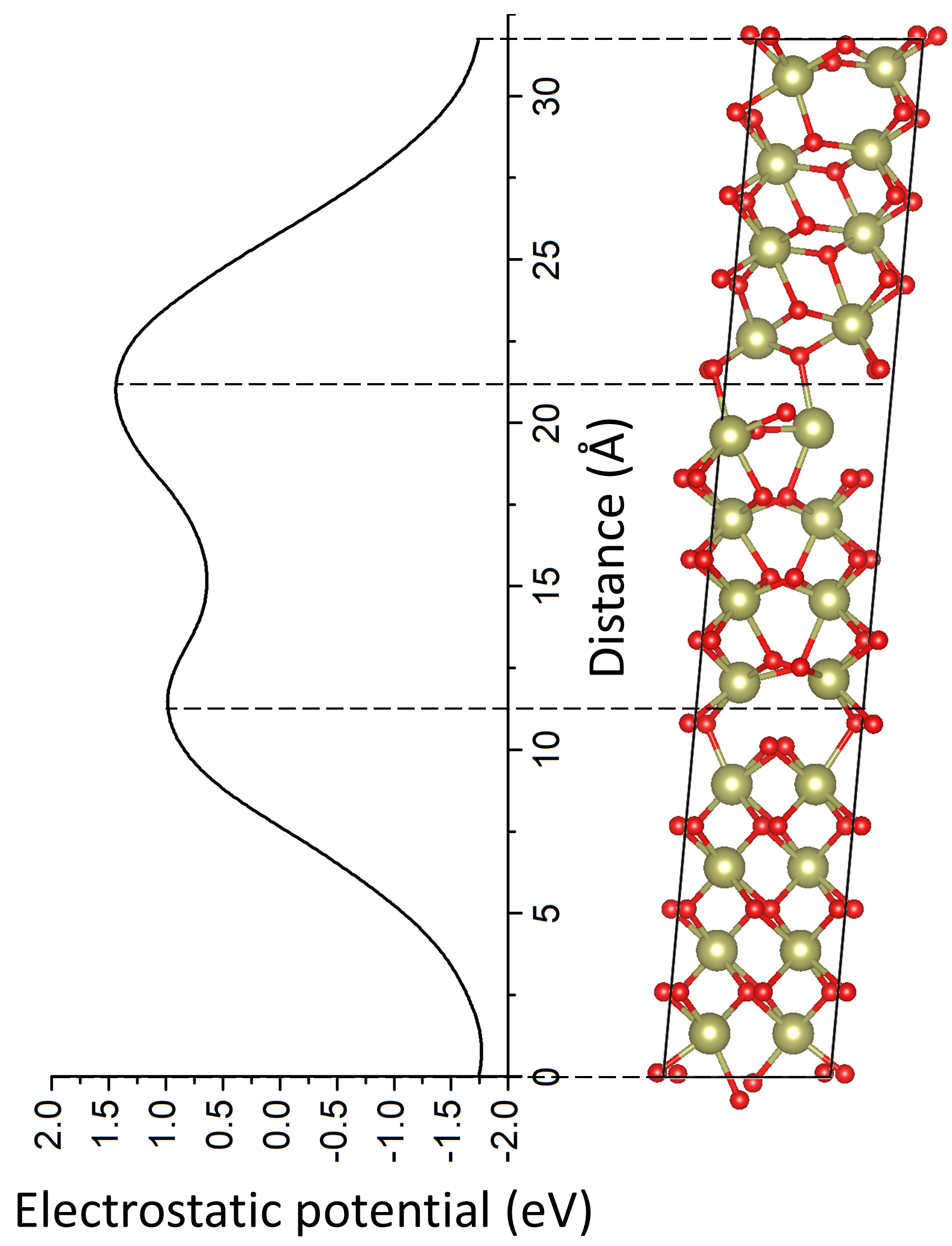}
    \caption{Results for the ``tricolor'' tetragonal/oIII-1/monoclinic
      superlattice. The structure and the macroscopically-averaged
      electrostatic potential are shown.}
    \label{fig:tricolor}
\end{figure}

In view of these results, one can pose the following provocative
question: What happens if we consider a tricolor superlattice of
sorts, formed by tetragonal, oIII-1, and monoclinic layers?
Figure~\ref{fig:tricolor} summarizes our results for this case,
showing that both tetragonal/oIII-1 and oIII-1/monoclinic interfaces
act as traps for holes. Importantly, this result is consistent with
what we obtain for the corresponding interfaces in the bicolor
superlattices -- i.e., the tetragonal/oIII-1 interface around
Hf-plane~8 in Fig.~\ref{fig:t-oIII}(a) and the oIII-1/monoclinic
interface around Hf-plane~4 in Fig.~\ref{fig:m-oIII}(a),
respectively. Thus, subject to these boundary conditions, the oIII-1
layer displays a positive depolarizing field in its upper half, which
turns negative in its lower half. This would seem to suggest that we
have a head-to-head ferroelectric domain wall in the middle of the
layer; note, though, that the polar distortion -- i.e., the oIII-1
structure -- remains the same throughout.

\begin{figure}
    \centering
    \includegraphics[width=0.8\linewidth]{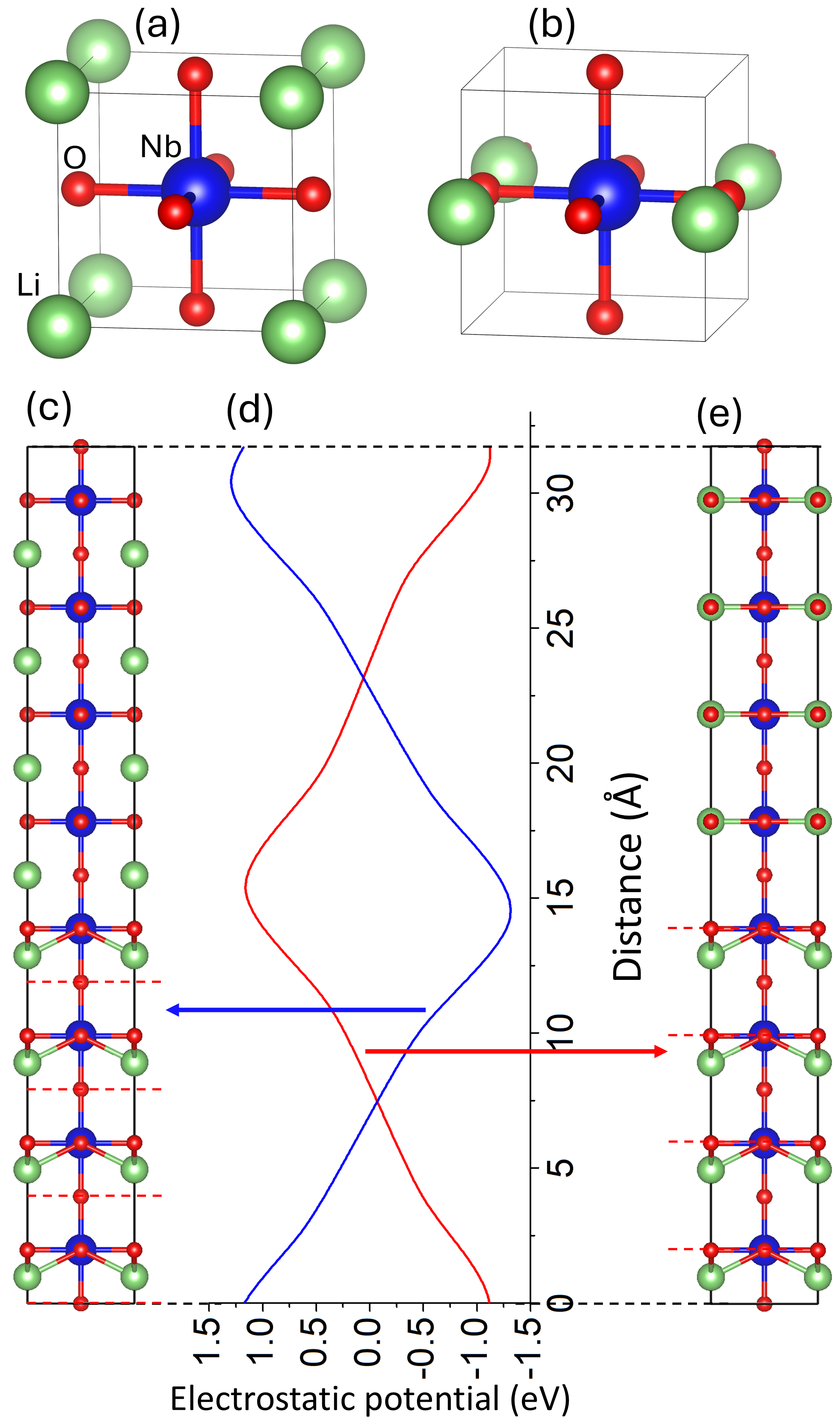}
    \caption{Sketches of LiNbO$_{3}$ in a bulk-like cubic perovskite
      structure (a) and in an alternative centrosymmetric structure
      (b). Panels~(c) and (e) show superlattices composed of polar and
      non-polar layers. The non-polar layers correspond to the
      structures in panels~(a) and (b), respectively. The polar layer
      is identical in both superlattices. In panel~(c), the polar
      layer can be seen as obtained from the cubic perovskite lattice
      by moving Li atoms up, by a quarter of a cell, with respect to
      cubic-like reference positions marked by dashed red lines. In
      panel~(e), the polar layer can be seen as obtained from the
      hypothetical structure in panel~(b) by moving Li atoms down by a
      quarter of a cell. Panel~(d) shows the computed
      macroscopically-averaged electrostatic potential for these two
      superlattices.}
    \label{fig:LNO}
\end{figure}

At this point, another natural question emerges: Are these effects
restricted to a particular ferroelectric class, of which hafnia would
be a representative, or are they general? We have not been able to
find any clear-cut criterion -- e.g., based on symmetry -- to qualify
the ferroelectric compounds whose polarization will be conditioned by
such extrinsic factors. Rather, we believe that this question boils
down to the practical aspects captured in Fig.~\ref{fig:schematic},
namely, the existence of several inequivalent and accessible (i.e.,
physically-relevant) switching paths and reference states. To explore
this point, consider a material with the perovskite structure and
composition LiNbO$_{3}$. (We acknowledge LiNbO$_{3}$ is not a
perovskite in reality. The choice is convenient in this context,
though, as Li$^{1+}$ is a small cation that could conceivably diffuse
in the perovskite lattice while keeping the system insulating.) Let us
take the ideal cubic perovskite phase (Fig.~\ref{fig:LNO}(a)) as the
centrosymmetric configuration~B in Fig.~\ref{fig:schematic}. Consider
the superlattice sketched in Fig.~\ref{fig:LNO}(c), with one layer
fixed in configuration~B (cubic) and a second layer where Li is moved
upwards by 1/4 of the cubic lattice constant. We can associate the
resulting polar state to configuration~A in
Fig.~\ref{fig:schematic}. As shown in the corresponding results for
the electrostatic potential (blue line in Fig.~\ref{fig:LNO}(d)), the
upward movement of the Li cations creates a negative depolarizing
field; that is, it corresponds to a positive polarization. This is
compatible with our default expectations.

Let us consider now the hypothetical non-polar structure in
Fig.~\ref{fig:LNO}(b), where the Li cations lie within the NbO$_{2}$
plane. This structure can be viewed as corresponding to
configuration~C in Fig.~\ref{fig:schematic}. Consider the superlattice
in Fig.~\ref{fig:LNO}(e), where a centrosymmetric layer in such
configuration~C alternates with a polar layer in
configuration~A. Remarkably, in this case the polar layer experiences
a positive depolarizing field, which suggests that it is negatively
polarized. As highlighted in the figure, the opposite depolarizing
fields and polarizations observed in these two superlattices can be
traced back to the structure of the non-polar layer, which effectively
determines the relevant reference to compute polarization changes. We
have thus reproduced, for the perovskite structure, the same
phenomenology discussed above for fluorite hafnia.

This exercise is admittedly academic, because the structure in
Fig.~\ref{fig:LNO}(b) has a very high energy and will not be
accessible in practice. For our hypothetical perovskite-like
LiNbO$_{3}$, the configurations in Figs.~\ref{fig:LNO}(a) and
\ref{fig:LNO}(b) lie within 0.22~eV per formula unit (f.u.) of each
other, but are about 0.70~eV/f.u. above the non-perovskite ground
state of the material. For actual ferroelectric perovskites like
BaTiO$_{3}$ and PbTiO$_{3}$, the structure in Fig.~\ref{fig:LNO}(b) is
predicted to be metallic, lying 5.4~eV/f.u. and 3.2~eV/f.u. above the
cubic polymorph, respectively. We thus contend that, as a matter of
principle, perovskite and hafnia ferroelectrics are not different in
what pertains to the subtleties discussed here. In practice, though,
these considerations are likely to be relevant for hafnia and other
fluorites, while we do not expect them to play any role in
perovskites.

Throughout this work, we ran numerous sanity checks to confirm that
our basic results and conclusions are solid. The most relevant are
summarized in Supp. Note~1 and Supp. Figs.~7 and 8, where we
explicitly check that the oIII-2 state leads to reversed depolarizing
fields compared to oIII-1, and Supp. Note~2, where we present a
quantitative discussion of the consistency between the computed
depolarizing fields and the spontaneous polarization of the oIII-1
state. We have also explored the behaviors described here in
superlattices combining HfO$_{2}$ with other fluorites. Our results,
briefly summarized in Supp. Note~3 and Supp. Figs.~9 to 12, align with
our general picture; yet, they are not as clear-cut due to the large
band-offsets and other interfacial effects associated with the
chemical modulation in such systems. Further investigation of the
interplay between polar and chemical effects remains for future
work. Finally, one might wonder whether the two possible polarization
values of the oIII-1 state, which depend on which phase is used as a
reference, might be explained by considerations pertaining to the
quantum of polarization or Berry-phase branch~\cite{kingsmith93}. We
refer the interested reader to literature addressing this
point~\cite{choe21,qi22,aramberri23,wu23}.

In summary, we have shown that it is possible to have ferroelectric
materials where the orientation of the polarization cannot be
ascertained from a mere inspection of the structure, but it rather
depends on the surrounding compounds. We have demonstrated that this
is the case in hafnia fluorite ferroelectrics, which feature multiple
centrosymmetric states that are physically relevant, that is,
accessible in typical samples. These considerations probably apply to
other ferroelectric families (e.g.,
CuInP$_{2}$S$_{6}$~\cite{brehm20,seleznev23}) where the multiplicity
of reference states translates into an intimate connection between
polarization switching and ionic diffusion. The extent to which these
features are reflected in the measured properties of hafnia
ferroelectrics -- and whether they can be leveraged in novel ways --
remains an open question. Opportunities for original concepts abound:
for example, our results suggest that hafnia's polarization can be
reversed indirectly, by acting on the surrounding materials, even if
hafnia itself remains in the same polar configuration. We thus hope
this work will stimulate further study and experimental confirmation
of these novel features, which expand the remarkably rich physics of
ferroelectricity.

{\sl Acknowledgements}.-- Work supported by the Luxembourg National
Research Fund though grant INTER/NOW/20/15079143/TRICOLOR. We thank
Hugo Aramberri (LIST) for fruitful discussions.

{\sl Methods}.-- We use density functional theory as implemented in
the Vienna ab-initio simulation package
(VASP)~\cite{kresse96,kresse99}. The Perdew-Burke-Ernzerhof (PBE) form
of the generalized gradient approximation (GGA) with the so-called
``PBEsol'' modification is employed to approximate the
exchange-correlation functional~\cite{perdew08}. The interaction
between core and valence electrons is treated within the
projector-augmented wave approach~\cite{blochl94}, considering the
following valence states explicitly: O's 2$s$ and 2$p$; Hf's 5$p$,
5$d$, and 6$s$; Zr's 4$s$, 4$p$, 4$d$, and 5$s$; Ce's 4$f$, 5$s$,
5$p$, 5$d$, and 6$s$;Li's 1$s$ and 2$s$; Nb's 4$s$, 4$p$, 4$d$, and
5$s$; Ba's 5$s$, 5$p$, and 6$s$; Pb's 5$d$, 6$s$, and 6$p$; Ti's 3$s$,
3$p$, 3$d$, and 4$s$. A plane-wave basis cut-off at 600~eV is used to
describe the electronic wave functions. A Gaussian smearing with a
width of 0.1~eV is used. For the HfO$_{2}$ superlattices, we sample
the Brillouin zone with a $4\times 4\times 1$ $k$-point mesh. The
interfacial ions between superlattice layers are allowed to relax,
with the Hellmann-Feynman forces converged below 0.01~eV/\AA, while
the ions in the middle of the layers are held fixed to their
(relative) positions in the corresponding polymorphs. The
macroscopically-averaged local electrostatic potential is then
computed for these partially relaxed structures. For LiNbO$_{3}$,
BaTiO$_{3}$, and PbTiO$_{3}$, the bulk cubic and bulk modified
structures are fully relaxed; here we employ a $k$-point grid of
$8\times 8\times 8$ for the calculations with a 5-atom perovskite
cell, and equivalent grids for the superlattices. The VASPKIT software
\cite{wang2021vaspkit} is used for postprocessing and the VESTA
software \cite{vesta11} to visualize structures.

\end{document}